\documentclass[pra,twocolumn,a4paper,superscriptaddress,aps,longbibliography]{revtex4-1}
\usepackage{epsf}
\usepackage{epsfig}
\usepackage{graphicx}
\usepackage{bm}
\usepackage[T1]{fontenc}
\usepackage[latin1]{inputenc}
\usepackage{times}
\usepackage{ulem}
\usepackage{color}
\usepackage[squaren,Gray]{SIunits}
\newcommand{\cora}[1]{\textcolor{black}{#1}}
\newcommand{\cors}[1]{\textcolor{black}{#1}}
\usepackage[pdfstartview=FitV,bookmarks=true,
            linktocpage=true,colorlinks=true,
            urlcolor=blue,linkcolor=blue,
            citecolor=blue]{hyperref}
\begin{document}

\title{{Particle-like wave {packets} in complex scattering systems}}

\author{Benoît Gérardin}
\affiliation{ESPCI Paris, PSL Research University, CNRS, Institut Langevin, UMR 7587, 1 rue Jussieu, F-75005 Paris, France}
\author{Jérôme Laurent}
\affiliation{ESPCI Paris, PSL Research University, CNRS, Institut Langevin, UMR 7587, 1 rue Jussieu, F-75005 Paris, France}
\author{Philipp Ambichl}
\affiliation{Institute for Theoretical Physics, Vienna University of Technology (TU Wien), Wiedner Hauptstraße 8-10/136, A-1040 Vienna, Austria}
\author{Claire Prada}
\affiliation{ESPCI Paris, PSL Research University, CNRS, Institut Langevin, UMR 7587, 1 rue Jussieu, F-75005 Paris, France}
\author{Stefan Rotter}
\affiliation{Institute for Theoretical Physics, Vienna University of Technology (TU Wien), Wiedner Hauptstraße 8-10/136, A-1040 Vienna, Austria}
\author{Alexandre Aubry}
\email{alexandre.aubry@espci.fr}
\affiliation{ESPCI Paris, PSL Research University, CNRS, Institut Langevin, UMR 7587, 1 rue Jussieu, F-75005 Paris, France}

\date{\today}

\begin{abstract}
A wave packet undergoes a strong spatial and temporal dispersion while propagating through a complex medium. This wave scattering is often seen as a nightmare in wave physics whether it be for focusing, imaging or communication purposes. Controlling wave propagation through complex systems is thus of fundamental interest in many areas, ranging from optics or acoustics to medical imaging or telecommunications. Here, we study the propagation of elastic waves in a cavity and a disordered waveguide by means of laser interferometry. \cora{From the direct experimental access to the time-delay matrix of these systems, we demonstrate the existence of particle-like wave packets that remain focused in time and space throughout their complex trajectory}.
%travel along individual classical trajectories. 
Due to their limited dispersion, \cora{their selective excitation} will be crucially relevant for all applications involving selective wave focusing and efficient information transfer through complex media.
\end{abstract}

\maketitle

\section{Introduction}

Waves propagating in complex media typically undergo diffraction and multiple scattering at all the inhomogeneities they encounter. As a consequence, a wave packet {suffers from} strong temporal and spatial dispersion while propagating through a scattering medium. {Eventually, the} incident wave is converted into a diffuse halo that gives rise to a {complicated} interference pattern {(speckle)} at the output of the medium. Albeit complex, this wave-field remains, however, deterministic. By actively shaping the wave-field at the input, one can manipulate the interference between all the scattering paths that the wave can follow. On the one hand, {this insight } has given rise to spectacular focusing schemes in which scattering enables - rather than impedes - wave focusing and pulse compression \cite{derode,tanter,lerosey,geo1,vellekoop2007focusing,popoff,aulbach,mccabe,katz}. On the other hand, it can lead to an optimized control of wave transport \cite{vellekoop2,pendry,choi,shi}. A designed wave-front can{, \textit{e.g.},} be completely transmitted/reflected at will \cite{choi3,gerardin} {as a result of a multiple scattering interference that is intrinsically narrowband \cite{hsu}. Here we will aim at the more challenging goal to generate states that are fully transmitted/reflected, yet very robust in a broadband spectral range. As we will demonstrate explicitly, this goal can be reached by way of highly collimated scattering states that are concentrated along individual particle-like bouncing patterns inside the medium \cite{rotter2011generating}. By avoiding the multi-path interference associated with conventional scattering states, \cora{these wave beams also avoid} the frequency sensitivity associated with this interference. 
\cora{As we shall see, particle-like scattering states give rise, in the time domain, to wave packets that remain focused in time and space throughout their trajectory within the medium. This crucial feature makes} these states uniquely suited for many applications in a variety of fields}, ranging from high intensity focused ultrasound \cite{jf,cochard} or underwater acoustics \cite{Edelmann,prada2007} to endoscopic microscopy \cite{Cizmar,salma,ploschner}, fibre optics \cite{Fan,Juarez,Carpenter,Xiong,Carpenter0} or telecommunications \cite{Salz,Raleigh,tulino}. 
 
{The key aspect of our experimental study is to demonstrate that these particle-like \cora{wave packets} can be created just based on the information stored in the scattering matrix \cite{rotter2011generating}. This highly dimension} $\mathbf{S}$-matrix relates any arbitrary wave-field at the input to the output of the scattering medium, and in principle, allows the reconstruction or prediction of either. It fully describes wave propagation across a scattering medium {and can meanwhile be routinely measured not only in acoustics \cite{sprik,aubry}, but also in microwave technology \cite{shi,Dietz} and optics \cite{popoff,choi}.} {The sub-blocks of the scattering matrix contain the complex-valued} transmission {($\mathbf{t},\mathbf{t'}$)} and reflection {($\mathbf{r}$, $\mathbf{r'}$)} matrices with a certain number $N$ of input and output channels,
\begin{equation}
\mathbf{S}= \left( \begin{array}{cc}
\mathbf{r} & \mathbf{t'} \\
\mathbf{t} & \mathbf{r'}
\end{array} 
 \right).
 \label{eq:S_blocks}
\end{equation}
{To describe the statistical properties of $\mathbf{S}$ for wave transport through complex media, random matrix theory (RMT) has been very successful \cite{beenakker}}. One of the striking result of RMT is the universal bimodal distribution followed by the transmission eigenvalues $T$ of $\mathbf{tt^{\dag}}$ \cite{footnote1} through diffusive media \cite{beenakker,dorokhov,imry} or chaotic cavities \cite{baranger1994mesoscopic,jalabert1994universal}. In contradiction with {a} classical diffusion or chaotic picture, {a substantial fraction of} propagation channels are found to be essentially closed ($T\sim0$) or open ($T\sim1$). {Going beyond such a statistical approach}, Rotter et al. \cite{rotter2011generating} {recently} showed how a {system-specific} combination of {fully open or fully closed} channels may lead to scattering states that follow the particle-like bouncing pattern of a classical trajectory throughout the entire scattering process. Such particle-like scattering states {with transmission close to 1 or 0 are eigenstates} of the Wigner-Smith time-delay matrix: 
\begin{equation}
\label{Q}
\mathbf{Q}= - \frac{i}{2\pi} \mathbf{S}^{\dag} \partial_f \mathbf{S},
\end{equation}
where $\partial_f$ denotes {the} derivative {with respect to} the frequency $f$. Originally introduced by Wigner in nuclear scattering theory \cite{wigner1955lower} and extended by Smith to multichannel scattering problems \cite{smith1960lifetime}, the $\mathbf{Q}$-matrix generally describes the time {that the incoming wave accumulates due to the} scattering process: each eigenvalue yields {the time delay of the associated scattering state. Compared to a mere study of the $\mathbf{S}$-matrix, the $\mathbf{Q}$-matrix provides an elegant and powerful tool to harness the dispersion properties of a complex medium.} In this article, we show, in particular, how {a} time-delay eigenstate {can be engineered to be ``particle-like'' not only in its stationary wave function patterns \cora{\cite{rotter2011generating}}, but also in the sense that a {non-dispersive} wave packet can be propagated along the corresponding particle-like bouncing pattern.} The associated eigenvalue of $\mathbf{Q}$ {then} corresponds to the {propagation-time of this wave packet}.

\begin{figure}[htbp]
\center
\includegraphics[width=8.5cm]{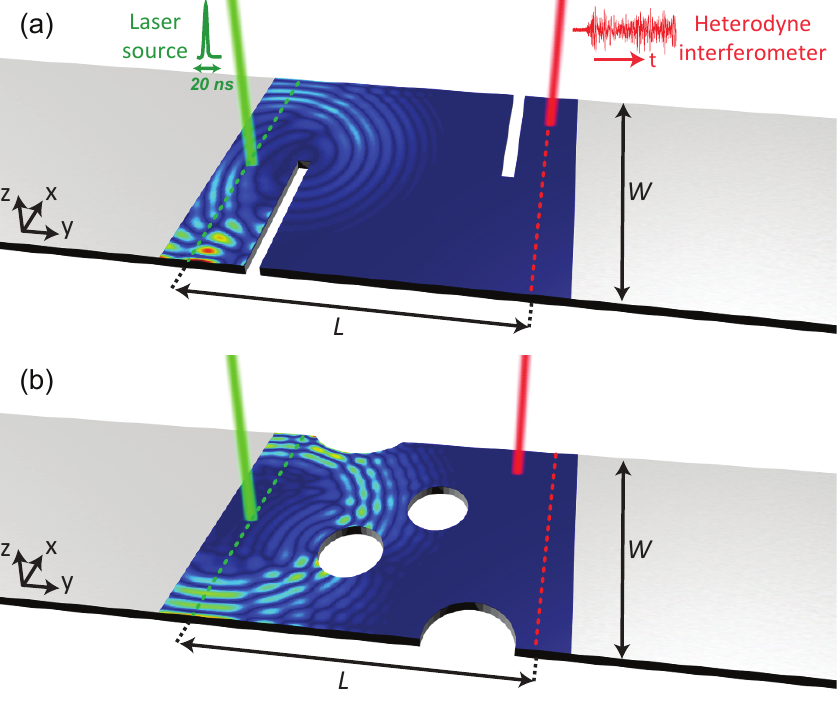}
\caption{The two systems under investigation consist ({a}) of a regular cavity and (b) of a disordered slab machined in an elastic plate. In both configurations, the $\mathbf{S}$-matrix is measured in the time-domain between two arrays of points placed on the left and right sides of the system (see Appendix \ref{appA}). Flexural waves are generated on each point by a pulsed laser \textit{via} thermo-elastic conversion over a focal spot of 1 mm$^2$. The normal component of the plate vibration is measured with an interferometric optical probe. The laser source and the probe are both mounted on 2D translation stages.}
\label{fig1}
\end{figure}

{Our experimental setup consists of an} elastic cavity and a disordered elastic wave guide at ultrasound frequencies [see Fig.~\ref{fig1}]. {In a first step we measure the entries of the} $\mathbf{S}$- and $\mathbf{Q}$-matrices over a large bandwidth using laser-ultrasonic techniques. The eigenvalues of the transmission matrix are shown to follow the expected bimodal distributions in both configurations. The wave-fields associated with the open/closed channels are monitored within each system in the time domain by laser interferometry. Not surprisingly, they are shown to be strongly dispersive as they combine various path trajectories {and thus many interfering scattering phases. To reduce this dispersion and to lift the degeneracy among the open/closed channels, we \cora{consider the} eigenstates of the $\mathbf{Q}$-matrix that have a well-defined time-delay, corresponding to a wave that follows a single path trajectory.} In transmission, a one-to-one association is found between time-delay eigenstates and ray-path trajectories. \cora{The corresponding wave functions are imaged in the time domain by laser interferometry.}
%The broadband generation of these states leads to spatio-temporal wave-packets whose propagation can be visualized in the time domain. 
The synthesized wave {packets} are shown to follow particle-like trajectories along which the temporal {spreading} of the incident pulse is {minimal}. In reflection, the $\mathbf{Q}$-matrix yields the {collimated} wave-fronts that focus selectively on each scatterer of a multi-target medium. Contrary to {alternative} approach{es based on time-reversal techniques} \cite{prada,prada2,popoff3,badon}, the discrimination between several targets is not based on their reflectivity but on their position. The eigenvalues of $\mathbf{Q}$ directly yield the time-of-flight of the pulsed echoes reflected by each scatterer.

\section{Experimental results}

\subsection{Revealing the open and closed channels in a cavity}

{The waves we excite and measure are flexural waves in} a duralumin plate of dimension $\unit{500\times40\times0.5}{\milli\cubicmetre}$ [see Fig.~\ref{fig1}].The frequency range of interest spans from \unit{0.23} to \unit{0.37}{\mega\hertz} ($\Delta f= 0.14$ MHz) {with a corresponding average wavelength $\lambda$ of} \unit{3.5}{\milli\metre}. {We thus have} access to $N = 2W/\lambda \sim 22$ independent channels, $W$ being the width of the elastic plate. Two complex scattering systems are built from the homogeneous plate: {(i) a regular cavity formed by cutting the plate over \unit{20}{\milli\metre} on both sides of the plate [see Fig.~\ref{fig1}(a)] and (ii) a scattering slab obtained by drilling several circular holes in the plate [see Fig.~\ref{fig1}(b)]. }
The thickness $L$ of each system is $\unit{45}{\milli\metre}$ and $\unit{52}{\milli\metre}$, respectively.

\begin{figure}[htbp]
\center
\includegraphics[width=7cm]{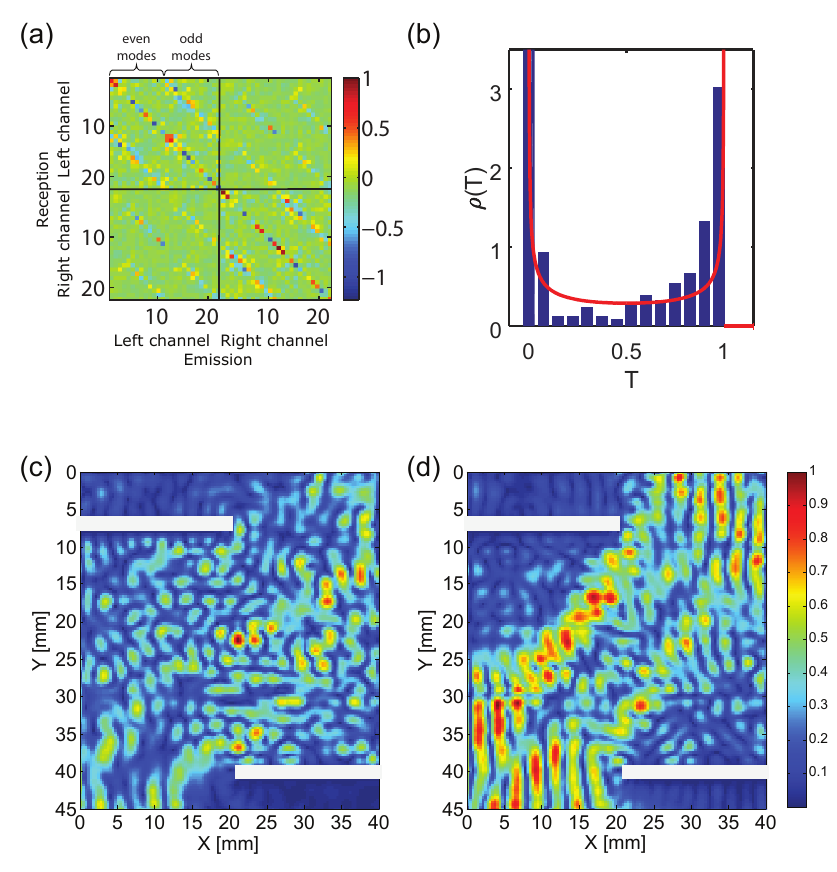}
\caption{({a}), Real part of the $\mathbf{S}$-matrix measured in the cavity [Fig.~\ref{fig1}(a)] at $f_0 = \unit{0.30}{\mega\hertz}$. The black lines delimit transmission and reflection matrices as depicted in Eq.~(\ref{eq:S_blocks}). ({b}), Transmission eigenvalue histogram, $\rho\left(T\right)$, averaged over the frequency bandwidth \cora{$f=0.23-0.37$} MHz. The distribution is compared to the bimodal law $\rho_b\left(T\right)$ [red continuous line, Eq.~(\ref{eq:bimodal})]. ({c})-({d}), Absolute values of the monochromatic wave-fields ($f_0 = \unit{0.30}{\mega\hertz}$) associated with the two {most} open channels.}
\label{fig2}
\end{figure}
The $\mathbf{S}$-matrix is measured for each system with the laser-ultrasonic set-up described in Fig.~\ref{fig1}, following the procedure explained in the Appendix~\ref{appA}. Transmission and reflection matrices are expressed in the basis of the modes of the homogeneous plate \citep{gerardin}. These eigenmodes and their eigenfrequencies have been determined theoretically using the thin elastic plate theory\citep{cross2001elastic,santamore2002surface}. They are renormalized such that each of them carries unit energy flux across the plate section\citep{gerardin}. Figure~\ref{fig2}(a) displays an example of an $\mathbf{S}$-matrix recorded at the central frequency $f_0=\unit{0.30}{\mega\hertz}$ for the cavity. Most of the energy emerges along the main diagonal and two sub-diagonals \cite{footnote2} of the reflection/transmission matrices. These reflection and transmission matrix elements correspond to specular reflection of each mode on the cavity boundaries and to the ballistic transmission of the incident wave-front, respectively.

We first focus on the statistics of the transmission eigenvalues $T_l$ computed from the measured $\mathbf{t}$-matrix (see Appendix \ref{appB}). Their distribution, $\rho(T)$ is estimated by averaging the corresponding histograms over the frequency bandwidth. Figure~\ref{fig2}(b) shows the comparison between the distribution {measured in the rectangular cavity} and the bimodal law $\rho_b$ which is theoretically expected in the chaotic regime\cite{baranger1994mesoscopic,jalabert1994universal},
\begin{equation}
\rho_b\left(T\right) = \frac{1}{\pi\sqrt{T\left(1-T\right)}}
\label{eq:bimodal}
\end{equation}
Even though our system is not {chaotic, but exhibits regular dynamics}, a good agreement is found between the measured eigenvalue distribution and the RMT predictions, confirming previous numerical studies \cite{aigner2005shot}. A similar bimodal distribution of transmission eigenvalues is obtained in the disordered plate, as shown in the Supplemental Material \cite{supp}.

Whereas the eigenvalues $T_l$ of $\mathbf{tt^{\dag}}$ yield the transmission coefficients of each eigenchannel, the corresponding eigenvectors provide the combination of incident modes that allow to excite this specific channel. Hence, the wave-field associated with each eigenchannel can be measured by propagating the corresponding eigenvector. {To that aim, the whole system is scanned with the interferometric optical probe. A set of impulse responses is measured between the line of sources at the input and a grid of points that maps the medium. The wave function associated with a scattering state is then deduced by a coherent superposition of these responses weighted by the amplitude and phase of the eigenvector at the input (see Appendix \ref{appC}). Hence all the wave functions displayed in this article are only composed of experimentally measured data and do not imply any \cora{theoretical calculation or} numerical simulation}.

\begin{figure}[htbp]
\center 
\includegraphics[width=8cm]{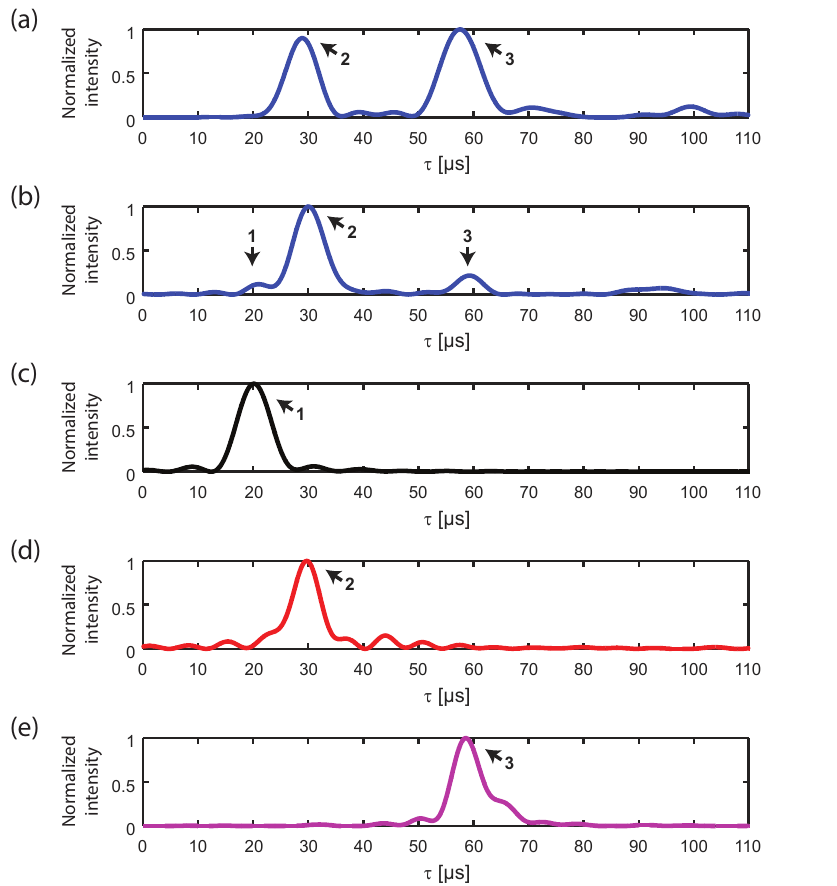}
\caption{The time trace of a scattering or time-delay eigenstates is computed from the set of $\mathbf{t}$-matrices measured over \cora{the frequency range $f=0.23-0
.37$ MHz} (see Appendix \ref{appB}). The output intensity is displayed versus time for, (a)-(b), the two first open channels displayed in Figs.~\ref{fig2}(c) and \ref{fig2}(d) and for, (c)-(e), the three time-delay eigenstates displayed in Fig.~\ref{fig3}. For each time trace, the different echoes are labeled with a number 1, 2 or 3 that corresponds respectively, to the direct, double or quadruple scattering paths highlighted in Fig.~\ref{fig3}. }
\label{fig2b}
\end{figure}

Figures \ref{fig2}(c) and \ref{fig2}(d) display the wave-field associated with the two {most} open eigenchannels ($T_l \sim 1$) of the cavity. Although such open channels allow a full transmission of the incident energy, they do not show a clear correspondence with a particular path trajectory. The same observation holds in the disordered plate \cite{supp}. As a consequence, the associated {scattering state undergoes multiple} scattering {when passing} through the cavity. Figures~\ref{fig2b}(a) and \ref{fig2b}(b) illustrate this dispersion by displaying the output temporal signal associated with the two open channels shown in Figs.~\ref{fig2}(c) and \ref{fig2}(d) (see Appendix \ref{appB}). Both signals contain several peaks occurring {at altogether three} different times of flight. As we will see further, each peak is {associated with a particular} path trajectory and can be addressed independently by means of the {Wigner-Smith time-delay} matrix.

\subsection{Addressing particle-like scattering states in a cavity}
The {Wigner-Smith time-delay} matrix $\mathbf{Q}$ is now investigated to generate coherent scattering states from the set of open channels. {Since $\mathbf{Q}$ is Hermitian when derived from a unitary $\mathbf{S}$-matrix} [Eq.~(\ref{Q})], the time-delay eigenstates $\mathbf{q}^{in}_m$ form an orthogonal and complete set of states, to each of which a real proper delay time $\tau_m$ can be assigned, such that $\mathbf{Q}\mathbf{q}^{in}_m=\tau_m\mathbf{q}^{in}_m$. In general, $\mathbf{q}^{in}_m$ is a $2N$-dimensional eigenvector which implies an injection from both the left and the right leads of the system. However, among this set of time-delay eigenstates, a subset features an incoming flux from {only} one lead that {also} exits through just one of the leads. {These are exactly the desired states that} belong to the subspace of open or closed channels and display trajectory-like wave function patterns. 
\begin{figure}[htbp]
\center
\includegraphics[width=7cm]{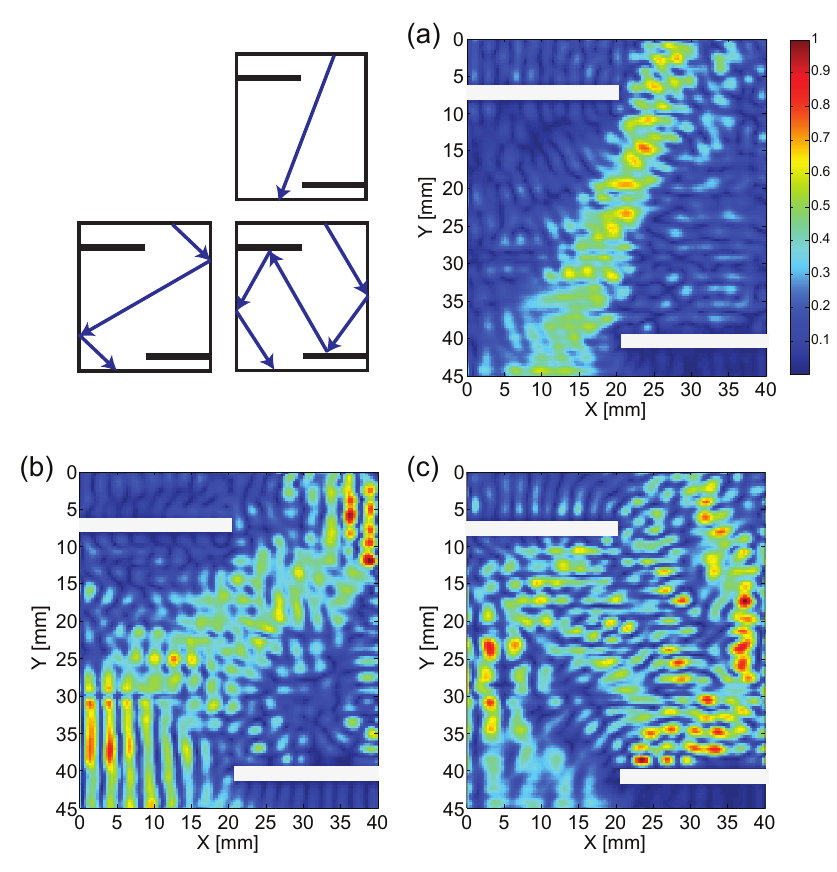}
\caption{Absolute value of the wave-field associated with the three particle-like scattering states derived from the matrix $\mathbf{Q}$. The corresponding time delays, (a), $\tau_m = \unit{20}{\micro\second}$, (b), $\tau_m = \unit{30}{\micro\second}$ and, (c), $\tau_m = \unit{59}{\micro\second}$, {nicely match with the run-times of the corresponding classical trajectories shown in the inset (top left)}.}
\label{fig3}
\end{figure}

\begin{figure*}[htbp]
\center
\includegraphics[width=14cm]{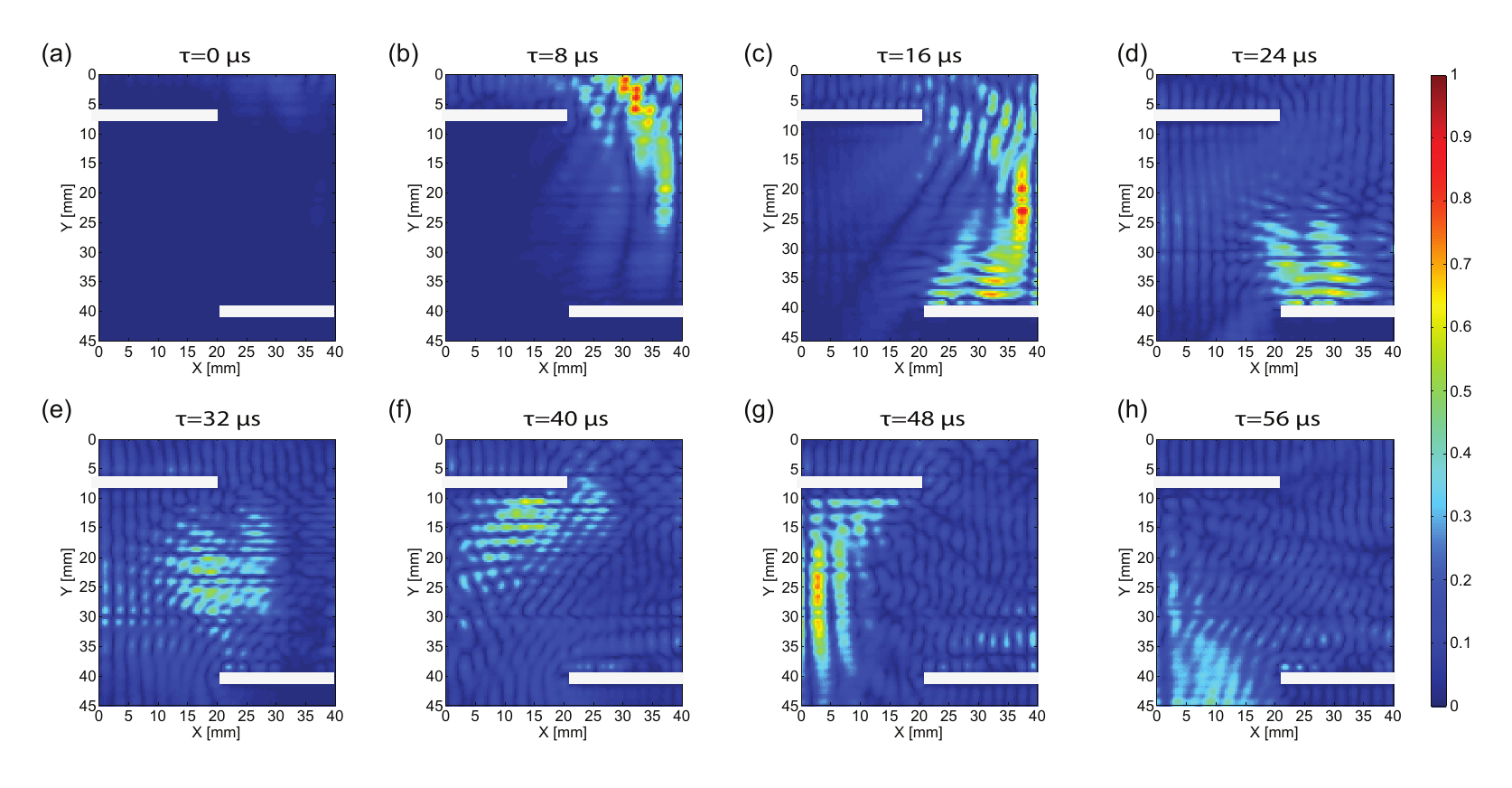}
\caption{A spatio-temporal wave-packet is synthesized from the particle-like scattering state displayed in Fig.~\ref{fig3}(c) \cora{over the frequency bandwidth $0.2-0.4$ MHz}. The different subsets (a)-(h) display successive snapshots of its propagation across the cavity versus time.}
\label{fig4}
\end{figure*}
{As was shown by Rotter \text{et al.} \cite{rotter2011generating}, the above arguments can be translated into a straightforward operational procedure (see Appendix \ref{appB}), which we apply here to identify} the particle-like scattering states among the time-delay eigenstates of the measured $\mathbf{Q}$-matrix. {The litmus test for this procedure in the present context will be to show that the three different time traces that are identifiable in the open transmission channels [see Figs.~\ref{fig2b}(a) and \ref{fig2b}(b)] can now be individually addressed through an associated particle-like state. The results we obtain for the cavity geometry [Fig.~\ref{fig1}(a)] fully confirm our first successful implementation of particle-like scattering states: The propagation of the states we obtain from our procedure yields monochromatic wave states that are clearly concentrated on individual bouncing patterns [see Fig.~\ref{fig3}].} Whereas Fig.~\ref{fig3}(a) corresponds to the direct path between the input and output leads, Figs.~\ref{fig3}(b) and \ref{fig3}(c) display a more complex trajectory with two and four reflections on the boundaries of the cavity, respectively. The associated time-delays $\tau_m$ do correspond to the {run-}times of a particle that would follow the same trajectory at the group velocity $v_{g}\sim \unit{2.6}{\milli\meter\per\micro\second}$ of the flexural wave \cite{royer}. \cora{Their transmission coefficients $|t_m|$ are equal to $0.90$, $0.95$ and $0.85$, respectively, meaning that they are almost fully transmitted through the cavity.}

The time trace associated with each particle-like scattering state is computed from the frequency-dependent $\mathbf{t}$-matrix (see Appendix \ref{appB}). The result is displayed in Figs.~\ref{fig2b}(c), \ref{fig2b}(d) and \ref{fig2b}(e). {Unlike the open transmission channels studied above, each particle-like state gives rise to} a single pulse that arrives at the output temporally unscattered at time $\tau=\tau_m$. Figure~\ref{fig2b} also shows that each temporal peak in the time trace of the open channels can be attributed to a particular path trajectory. We may thus conclude that the open channel displayed in Fig.~\ref{fig2}(c) is mainly associated with the double and quadruple scattering paths displayed in Figs.~\ref{fig3}(b) and \ref{fig3}(c). The open channel displayed in Fig.~\ref{fig2}(d) consists of a linear combination of the paths displayed in Fig.\ref{fig3}. {This association is also confirmed by explicitly analyzing the vectorial decomposition of the particle-like state in terms of the transmission eigenchannel basis.}

\cora{The frequency dependence of the particle-like scattering states is investigated in the Supplemental Material \cite{supp}. They are shown to be stable over the frequency ranges $f=0.2-0.6$ MHz [Fig.~\ref{fig3}(a)], $f=0.3-0.6$ MHz [Fig.~\ref{fig3}(b)] and $f=0.2-0.4$ MHz [Fig.~\ref{fig3}(c)]. The corresponding bandwidths are at least one order of magnitude larger than the frequency correlation \cors{width} of the transmission matrix coefficients which is equal to 0.02 MHz \cite{supp}. This proves the robustness of particle-like states over a broadband spectral range. Given this non-dispersive feature,} they turn out to be perfect candidates {also for the formation of minimally dispersive wave packets in the time domain}. To check this conjecture, {we investigate here the spatio-temporal} wave functions of these states \cora{over the aforementioned bandwidths} (see Appendix \ref{appB}). The propagation of the particle-like wave-packets through the cavity can be visualized in the three first movies of the Supplemental Material \cite{supp}. Figure~\ref{fig4} displays successive snapshots of the wave-packet synthesized from the particle-like scattering state displayed in Fig.~\ref{fig3}(c). {Quite remarkably, the spatio-temporal focus} of the incident wave-packet is maintained throughout its trajectory despite the multiple scattering events it undergoes in the cavity.

\subsection{Lifting the degeneracy of {particle-like} scattering states}
\begin{figure*}[htbp]
\center 
\includegraphics[width=14cm]{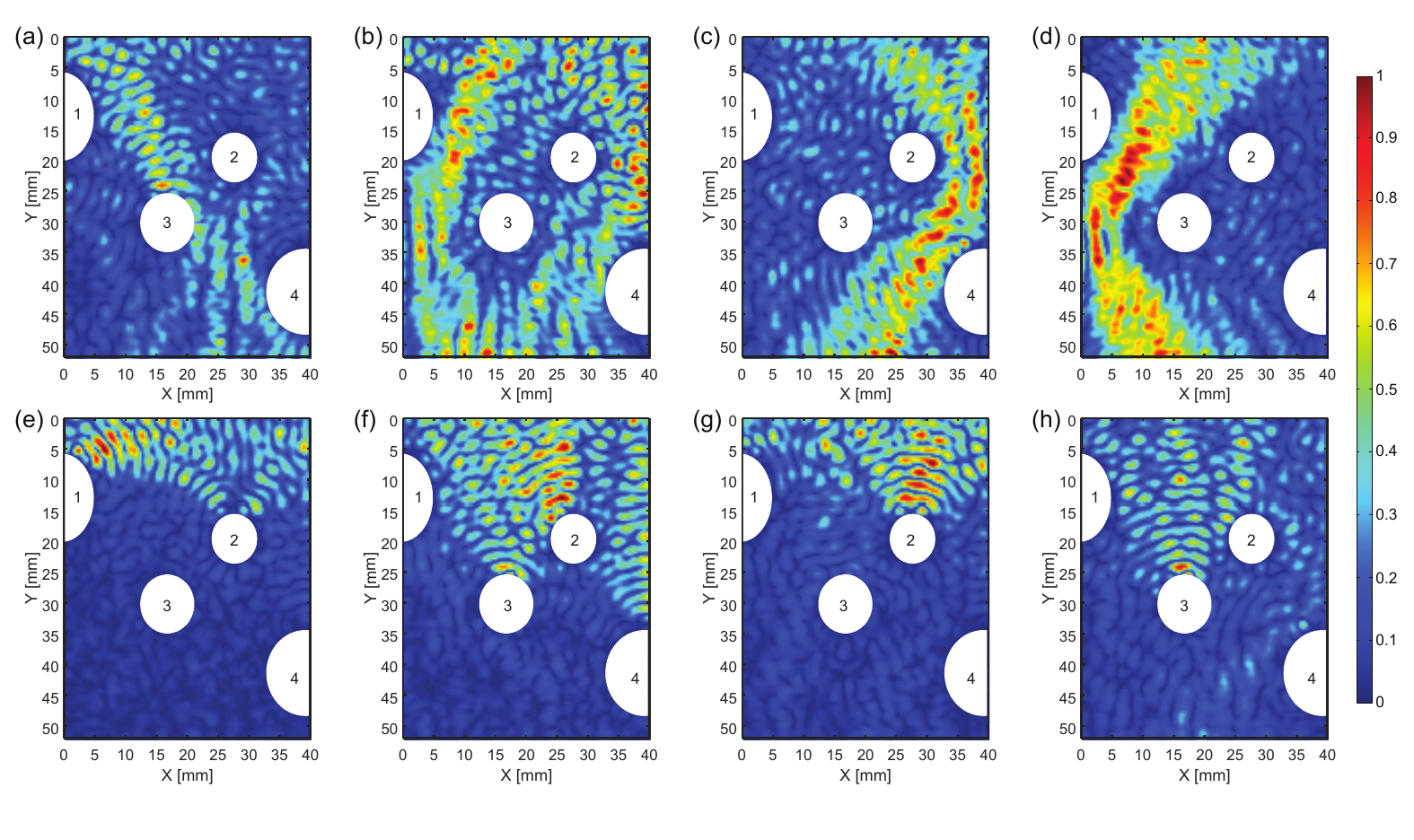}
\caption{Absolute value of the wave-field associated with several time-delay/scattering eigenstates of the disordered system at the central frequency $f_0$. 
(a), Transmitted particle-like scattering state ($\tau_m = \unit{29}{\micro\second}$\cora{, $|t_m|=0.84$}). (b) Degenerate time-delay eigenstate mixing two particle-like trajectories ($\tau_m = \unit{26}{\micro\second}$\cora{, $|t_m|=0.91$}). (c)-(d) Unmixing of the time-delay eigenstate displayed in (b) by considering different subspaces of the $\mathbf{S}$-matrix (see Appendix \ref{appD}). \cora{Their transmission coefficients $|t_m|$ are equal to 0.94 and 0.98, respectively}.  (e)-(f) Closed channels deduced from the $\mathbf{S}$-matrix ($T_l \sim 0$) showing simultaneous reflections from the scatterers of the disordered wave guide. \cora{({g}) Reflected time-delay eigenstate showing selective focusing on the scatterer n$^o$2 (time delay $\tau_m=\unit{11}{\micro\second}$, reflection coefficient $|r_m|=0.94$). (h) Reflected time-delay eigenstate showing selective focusing on the scatterer n$^o$3 (time delay $\tau_m=\unit{30}{\micro\second}$, reflection coefficient $|r_m|=0.88$).} }
\label{fig5}
\end{figure*}
{In a next step, we investigate particle-like} scattering states in the disordered wave guide [Fig.~\ref{fig1}(b)]. The corresponding $\mathbf{Q}$-matrix is measured at the central frequency $f_0=0.3$ MHz (see Appendix \ref{appA}). Figures~\ref{fig5}(a) and \ref{fig5}(b) display the monochromatic wave functions associated with two fully transmitted time-delay eigenstates. Whereas Fig.~\ref{fig5}(a) displays the typical features of a particle-like scattering state that winds its way through the scatterers, the time-delay eigenstate of Fig.~\ref{fig5}(b) is clearly associated with two scattering paths of identical length. {We thus encounter here a degeneracy in the time-delay eigenvalues that needs to be lifted by an additional criterion, such as by} considering {well-defined} subspaces of the measured $\mathbf{S}$-matrix \cite{andre}. In this instance, the two ray paths can be discriminated {by their different angles of incidence}. {Correspondingly, we consider two} subspaces $\mathbf{S'}$ of the original $\mathbf{S}$-matrix by keeping either positive or negative angles of incidence from the left lead (see Appendix \ref{appD}). The corresponding time-delay matrices lead to two distinct particle-like scattering states displayed in Figs.~\ref{fig5}(c) and \ref{fig5}(d). The two scattering paths that were previously mixed in the original time-delay eigenstate [Fig.~\ref{fig5}(b)] are now {clearly} separated. \cora{The frequency stability of these states is investigated in the Supplemental Material \cite{supp}. They are shown to be stable over the frequency ranges $f=0.2-0.4$ MHz [Fig.~\ref{fig5}(c)] and $f=0.2-0.5$ MHz [Fig.~\ref{fig5}(d)].} {The corresponding particle-like wave packets are shown in the two last movies of Supplemental Material \cite{supp} where the high quality of their focus in space and time is immediately apparent.}

\subsection{Revealing time-delay eigenstates in reflection.}
Time-delay eigenstates can also {result from a suitable} combination of closed channels. Figures~\ref{fig5}(e) and \ref{fig5}(f) display two {such} closed channels derived from the $\mathbf{S}$-matrix of the disordered slab. The closed channels combine multiple reflections from the holes of the scattering layer. {However,} the closed channel displayed in Fig.~\ref{fig5}(e) mixes the contributions from the scatterers labeled 1 and 2. {Also} the closed channel displayed in Fig.~\ref{fig5}(f) is associated with reflections {from altogether} three scatterers (2, 3 and 4). A simple eigenvalue decomposition of $\mathbf{rr^{\dag}}$ or $\mathbf{tt^{\dag}}$ does not allow a discrimination between the scatterers. On the contrary, the analysis of the $\mathbf{Q}$-matrix allows a one-to-one correspondence with each scatterer based on a time-of-flight discrimination. Figures~\ref{fig5}(g) and \ref{fig5}(h) actually display the wave functions associated with two reflected time-delay eigenstates. Each of these eigenstates is associated with a reflection from a single scatterer (2 and 3, respectively). The corresponding time delays $\tau_m$ given in the caption of Fig.~\ref{fig5} are directly related to the depth $z$ of each scatterer, such that $\tau_m \sim 2 z/v_g$.

\section{{Discussion}}

The first point we would like to {emphasize} is the relevance of the time-delay matrix for selective focusing and imaging in multi-target media. The state-of-the-art approach is the DORT method (French acronym for Decomposition of the Time Reversal Operator). Initially developed for ultrasound \cite{prada,prada2} and more recently extended to optics \cite{popoff3,badon}, this widely used approach takes advantage of the reflection matrix $\mathbf{r}$ to focus iteratively by time reversal processing on each scatterer of a multi-target medium. Mathematically, the time-reversal invariants can be deduced from the eigenvalue decomposition of the time reversal operator $\mathbf{r}\mathbf{r^{\dag}}$ or, equivalently, from the singular value decomposition of $\mathbf{r}$. On the one hand, the eigenvectors of $\mathbf{r}$ should, in principle, allow selective focusing and imaging of each scatterer. On the other hand, the associated eigenvalue directly yields the scatterer reflectivity. However, a one-to-one association between each eigenstate of $\mathbf{r}$ and each scatterer only exists under a single-scattering approximation and if the scatterers exhibit sufficiently different reflectivities. Figures~\ref{fig5}(e) and \ref{fig5}(f) illustrate this limit. Because the holes display similar scattering cross-sections in the disordered slab, closed channels are associated with several scatterers at once. On the contrary, the time-delay matrix allows a discrimination between scatterers based on the time-of-flight of the reflected echoes [see Figs.~\ref{fig5}(g) and \ref{fig5}(h)]. Moreover, unlike the DORT method, a time-delay analysis also allows to discriminate the single scattering paths from multiple scattering events, the latter ones corresponding to longer time-of-flights. Hence, the time-delay matrix provides an alternative and promising route for selective focusing and imaging in multi-target media.

{A second relevant point} to discuss is the nature of transmitted time-delay eigenstates in other complex systems. Recently, Carpenter \textit{et al.} \cite{Carpenter} and Xiong \textit{et al.}  \cite{Xiong} investigated the group delay operator, $i/(2\pi) \mathbf{t}^{-1} \partial_{f} \mathbf{t}$, in a multi-mode optical fiber. The eigenstates of this operator are known as the principal modes in fiber optics \cite{Fan,Juarez}. For a principal mode input, a pulse {that is sufficiently narrow-band} reaches the output temporally {un-distorted}, although it may have been strongly scattered and dispersed along the length of propagation. On the contrary, for a {particle-like} scattering state input, the focus of the pulse is not only retrieved at the output of the scattering medium but maintained throughout its {entire propagation through the system. This crucial difference provides particle-like states with a much broader frequency stability than principal modes, which translates into the possibility to send much shorter pulses through these particle-like scattering channels. Last but not least, we also emphasize that even though particle-like states are unlikely to occur in diffusive scattering media, the time-delay eigenstates are still very relevant also in such a strongly disordered context. Consider here, \textit{e.g.}, that the} eigenstates {with the longest time delay} can be of interest for energy storage, coherent absorption \cite{chong} or lasing \cite{bachelard,Bachelard2} purposes. From a more fundamental point-of-view, the trace of the $\mathbf{Q}$- matrix directly provides the density of states of the scattering medium \cite{davy}{, a quantity that turns out to be entirely independent of the mean free path in a disordered system \cite{pierrat}.}

\cora{Finally, we would like to stress the impact of our study on other fields of wave physics and its extension to more complex geometries. For this experimental proof-of-concept, a measurement of the wave-field inside the medium was required in order to image the wave functions and prove their particle-like feature. However, such a sophisticated protocol is not needed to physically address particle-like wave packets. A simple measurement of the scattering matrix (or a subpart of it) at neighboring frequencies \cite{Carpenter} yields the time-delay matrix from which particle-like state inputs can be extracted. Such a measurement can be routinely performed through 3D scattering media whether it be in optics \cite{popoff,choi}, in the microwave regime \cite{shi,Dietz} or in acoustics \cite{sprik,aubry}. As to the generation of particle-like \cors{wave} packets, multi-element technology is a powerful tool for the coherent control of acoustic waves and electromagnetic waves \cite{Mosk_review}. Moreover, recent progress in optical manipulation techniques now allows for a precise spatial and temporal control of light at the input of a complex medium \cite{Mosk_review}. Hence there is no obstacle for the experimental implementation of particle-like wave packets in other fields of wave physics.} \cors{At last, we would like to stress the fact that in our experimental implementation we are in the limit of only a few participating modes with a wavelength that is comparable to the spatial scales of the system. In this limit the implementation of particle-like states is truly non-trivial since interference and diffraction dominates the scattering process as a whole. When transferring the concept to the optical domain one may easily reach the geometric optics limit where the wavelength is much shorter than most spatial scales of the system and particle-like states may in fact be much easier to implement in corresponding complex optical media.}

\section{Conclusion}

In summary, {we experimentally implemented particle-like scattering states in complex scattering systems. Based on an experimentally determined time-delay matrix\cora{, we have demonstrated the existence of} wave packets that follow particle-like bouncing patterns in transmission through or in reflection from a complex scattering landscape. \cora{Strikingly, these wave packets have been shown to remain focused in time and space throughout their trajectory within the medium.} We are convinced that the superior properties of these states in terms of frequency stability and spatial focus will make them very attractive for} many applications of wave physics, ranging from focusing to imaging or communication purposes. {In transmission, the efficiency of these states in terms of information transfer as well as their focused in- and output profile will be relevant.} In reflection, selective focusing based on a time-of-flight discrimination will be a powerful tool to overcome aberration and multiple scattering in detection and imaging problems. 

\section{Acknowledgments}

The authors wish to thank A. Derode for fruitful discussions and advices. The authors are grateful for funding provided by LABEX WIFI (Laboratory of Excellence within the French Program Investments for the Future, ANR-10-LABX-24 and ANR-10-IDEX-0001-02 PSL*). B.G. acknowledges financial support from the French ``Direction Générale de l'Armement''(DGA). P.A. and S.R. are supported by the Austrian Science Fund (FWF) through Projects NextLite F49-10 and I 1142-N27 (GePartWave).

\begin{appendix}

\section{\label{appA}Experimental procedure}

The first step of the experiment consists in measuring the impulse responses between two arrays of points placed on the left and right sides of the disordered slab [see Fig.~\ref{fig1}]. These two arrays are placed 5 mm away from the disordered slab. The array pitch is {$0.8$ mm} (\textit{i.e} $< \lambda/2$) which guarantees a satisfying spatial sampling of the wave field. Flexural waves are generated in the thermoelastic regime by a pumped diode Nd:YAG laser (THALES Diva II) providing pulses having a \unit{20}{\nano\second}  duration  and  \unit{2.5}{\milli\joule} of  energy. The out-of-plane component of the local vibration of the plate is measured with a heterodyne interferometer. This probe is sensitive to the phase shift along the path of the optical probe beam. The calibration factor for mechanical displacement normal to the surface (\unit{100}{\milli \volt \per \nano \meter}) was constant over the detection bandwidth (\unit{100}{\kilo \hertz} - \unit{400}{\kilo \hertz}). Signals detected by the optical probe were fed into a digital sampling oscilloscope and transferred to a computer. The impulse responses between each point of the same array (left and right) form the time-dependent reflection matrices ($\mathbf{r}$ and $\mathbf {r'}$, respectively). The set of impulse responses between the two arrays yield the time-dependent transmission matrices $\mathbf{t}$ (from left to right) and $\mathbf{t'}$ (from right to left). From these four matrices, one can build the $\mathbf{S}$-matrix in a \textit{point-to-point} basis [Eq.~(\ref{eq:S_blocks})]. A discrete Fourier transform (DFT) of $\mathbf{S}$ is then performed over a time range $\Delta t=120$ $\mu$s that excludes the echoes due to reflections on the ends of the plate and ensures that most of the energy has escaped from the sample when the measurement is stopped. The next step of the experimental procedure consists in decomposing the $\mathbf{S}$-matrices in the basis of the flexural modes of the homogeneous plate. These eigenmodes and their eigenfrequencies have been determined theoretically using the thin elastic plate theory\citep{cross2001elastic,santamore2002surface,gerardin}. They are normalized such that each of them carries unit energy flux across the plate section. 
Theoretically, energy conservation would imply that $\mathbf{S}$ is unitary. In other words, its eigenvalues should be distributed along the unit circle in the complex plane.  However, as shown in a previous work \cite{gerardin}, this unitarity is not retrieved experimentally because of experimental noise. A dispersion of the eigenvalues $s_i$ of the $\mathbf{S-}$matrix is observed around the unit circle. We compensate for this undesirable effect by considering a normalized scattering matrix with the same eigenspaces but with normalized eigenvalues \cite{gerardin}. The $\mathbf{Q-}$matrix is then deduced from $\mathbf{S}$ using Eq.~(\ref{Q}). The frequency-derivative of $\mathbf{{S}}$ at $f=f_0$ is estimated from the centered finite difference,
$${\partial_f \mathbf{S}}\left(f_0\right)  = \frac{\mathbf{S}\left(f_0 +\delta f \right)- \mathbf{S}\left(f_0 -\delta f\right)}{2 \delta f},$$
with $\delta f = \unit{3}{\kilo \hertz}$.\\

\section{\label{appB} \cora{Revealing transmission/time-delay eigenchannels and their temporal/spectral features}}
The transmission and time-delay eigenchannels are derived from the matrices $\mathbf{S}$ and $\mathbf{Q}$ measured at the central frequency $f_0$. The transmission matrix $\mathbf{t}(f_0)$ (from the left to the right lead) is extracted from $\mathbf{S}(f_0)$ [Eq.~(\ref{eq:S_blocks})]. The output and input transmission eigenvectors, $\mathbf{u}_l$ and $\mathbf{v}_l$, are derived from the singular value decomposition of $\mathbf{t}(f_0)$:
$$\mathbf{t}(f_0)=\sum_l \sqrt{T_l(f_0)} \mathbf{u}_l(f_0) \mathbf{v}_l^{\dag}(f_0)  $$
with $T_l(f_0)$ the intensity transmission coefficient associated with the $l^{\mbox{th}}$ scattering eigenstate at the central frequency. The frequency-dependent amplitude transmission coefficient $t_l\left(f\right)$ of this eigenstate can be obtained from the set of transmission matrices $\mathbf{t}\left(f\right)$ measured over the whole frequency bandwidth, such that
$$
t_l\left(f\right) = \mathbf{u}^{\dag}_l(f_0)\mathbf{t}\left(f\right)\mathbf{v}_l(f_0)
$$
An inverse DFT of $t_l\left(f\right)$ finally yields the time-dependent amplitude transmission coefficient $t_l(\tau)$ of the $l^{\mbox{th}}$ scattering eigenstate measured at the central frequency $f_0$. The time traces displayed in Figs.~\ref{fig2b}(a) and \ref{fig2b}(b) correspond to the square norm of this quantity. 

The time-delay eigenchannels at the central frequency are derived from the eigenvalue decomposition of $\mathbf{Q}(f_0)$
$$\mathbf{Q}(f_0)=\sum_m {\tau_m} \mathbf{q}^{in}_m(f_0) \left [ \mathbf{q}^{in}_m(f_0) \right ]^{\dag}$$
The time-delay eigenvector $\mathbf{q}^{in}_m$ is a $2N$-dimensional column vector that can be decomposed as
$$\mathbf{q}^{in}(f_0)=\left( \begin{array}{c}
\mathbf{q}^{in}_{m,L}(f_0) \\
\mathbf{q}^{in}_{m,R}(f_0)
\end{array} 
 \right)$$
where $\mathbf{q}^{in}_{m,L}(f_0)$ and $\mathbf{q}^{in}_{m,R}(f_0)$ contain the complex coefficients of $\mathbf{q}^{in}_m(f_0)$ in the basis of the $N$ incoming modes in the left and right leads, respectively. Among the set of time-delay eigenstates, particle-like scattering states injected from the left lead should fulfill the following condition \cite{rotter2011generating}: 
$$|| \mathbf{q}_{m,L}^{in}(f_0) ||^2>>|| \mathbf{q}_{m,R}^{in}(f_0) ||^2 \simeq {0}.$$ 
A particle-like scattering state is thus associated with a $N$-dimensional input eigenvector $\mathbf{q}^{in}_{m,L}(f_0)$. The corresponding output eigenvector $\mathbf{q}^{out}_{m,R}(f_0)$ and transmission coefficient $t^{(q)}_m\left(f_0\right)$ can be deduced from the $\mathbf{t}$-matrix:
$$ \mathbf{t}(f_0) \mathbf{q}^{in}_{m,L}(f_0) =t^{(q)}_m\left(f_0\right) \mathbf{q}^{out}_{m,R}(f_0).$$
The frequency-dependent amplitude transmission coefficient $t_m^{(q)}\left(f\right)$ of this time-delay eigenstate can be obtained from the set of transmission matrices $\mathbf{t}\left(f\right)$ measured over the whole frequency bandwidth, such that
$$
t^{(q)}_m\left(f\right) = \left [ \mathbf{q}^{out}_{m,L}(f_0) \right ]^{\dag} \mathbf{t}\left(f\right) \mathbf{q}^{in}_{m,R}(f_0).
$$
An inverse DFT of $t^{(q)}_m\left(f\right)$ finally yields the time-dependent amplitude transmission coefficient $t^{(q)}_m(\tau)$ of the $m^{\mbox{th}}$ time-delay eigenstate measured at the central frequency $f_0$. The time traces displayed in Figs.~\ref{fig2b}(c), \ref{fig2b}(d) and \ref{fig2b}(e) correspond to the square norm of this quantity. \\

\section{\label{appC} \cora{Imaging spatio-temporal wave functions} of transmission/reflection and time-delay eigenchannels} 
Impulse responses are measured between the line of sources (denoted by the index $i$) and a grid of points (denoted by the index $j$) that maps the medium, following the same procedure as the one described above. The grid pitch is {1.3 mm}. This set of impulse responses forms a transmission matrix $\mathbf{k}(\tau)=[k_{ji}(\tau)]$. A discrete Fourier transform (DFT) of $\mathbf{k}(\tau)$ yields a set of frequency-dependent transmission matrices $\mathbf{k}(f)$. The lines of $\mathbf{k}(f)$ are then decomposed in the basis of the plate modes. The monochromatic wave-field $\mathbf{\Psi}(f)=[\psi_j(f)]$ associated with a transmission/reflection or time-delay eigenchannel is provided by the product between the matrix $\mathbf{k}(f)$ and the corresponding eigenvector ($\mathbf{v}_l(f_0)$ or $\mathbf{q}^{in}_{m,L}(f_0)$, respectively). The time-dependent wave-field $\mathbf{\Psi}(\tau)=[\psi_j(\tau)]$ is deduced by an inverse DFT over a frequency bandwidth \cora{of our choice}. Note that a {Hann} window function is priorly applied to $\mathbf{k}(f)$ to limit side lobes in the time-domain.

\section{\label{appD}Unmixing degenerated time-delay eigenstates.} 
Depending on the geometry of the scattering medium, the different scattering paths involved in a degenerated time-delay eigenstate can be discriminated either in the real space or in the spatial frequency domain \cite{andre}. Here, the time-delay eigenstate displayed in Fig.~\ref{fig5}(b) shows two scattering paths with opposite angles of incidence. Hence, they can be discriminated by analyzing each block of the $\mathbf{S}$-matrix in the spatial frequency domain. The left lead of each block is decomposed over the positive or negative angles of incidence. This sub-space of the $\mathbf{S}$-matrix, refered to as $\mathbf{S'}$, is then used to compute a reduced time-delay matrix $\mathbf{Q'}$ \cite{andre} such that 
$$\mathbf{Q'}=- \frac{i}{2\pi} \mathbf{S'}^{-1} \partial_f \mathbf{S'}. $$
Note that the transpose conjugate operation of Eq.~(\ref{Q}) is here replaced by an inversion of $\mathbf{S'}$ because of its non-unitarity \cite{andre}. Depending on the sign of the angle of incidence chosen for the left lead, the reduced matrix $\mathbf{Q'}$ provides the time-delay eigenstates displayed in Figs.~\ref{fig5}(c) and \ref{fig5}(d).

\end{appendix}

%\bibliography{PRX_Benoit} 

\renewcommand{\thefigure}{S\arabic{figure}}
\renewcommand{\theequation}{S\arabic{equation}}
\renewcommand{\thetable}{S\arabic{table}}
\setcounter{figure}{0} 
\setcounter{equation}{0} 
\setcounter{section}{0} 

\newpage

\begin{center}
\Large{\bf{Supplemental Material}}
\end{center}
\normalsize

This document provides further information on the scattering matrix measured in the disordered wave guide \cora{and the stability of particle-like states in the frequency domain. The captions of the Supplementary Movies are also listed at the end of the document. }

\section{Scattering matrix analysis in the disordered waveguide}
\begin{figure}[!ht]
\centering
\includegraphics[width=\columnwidth]{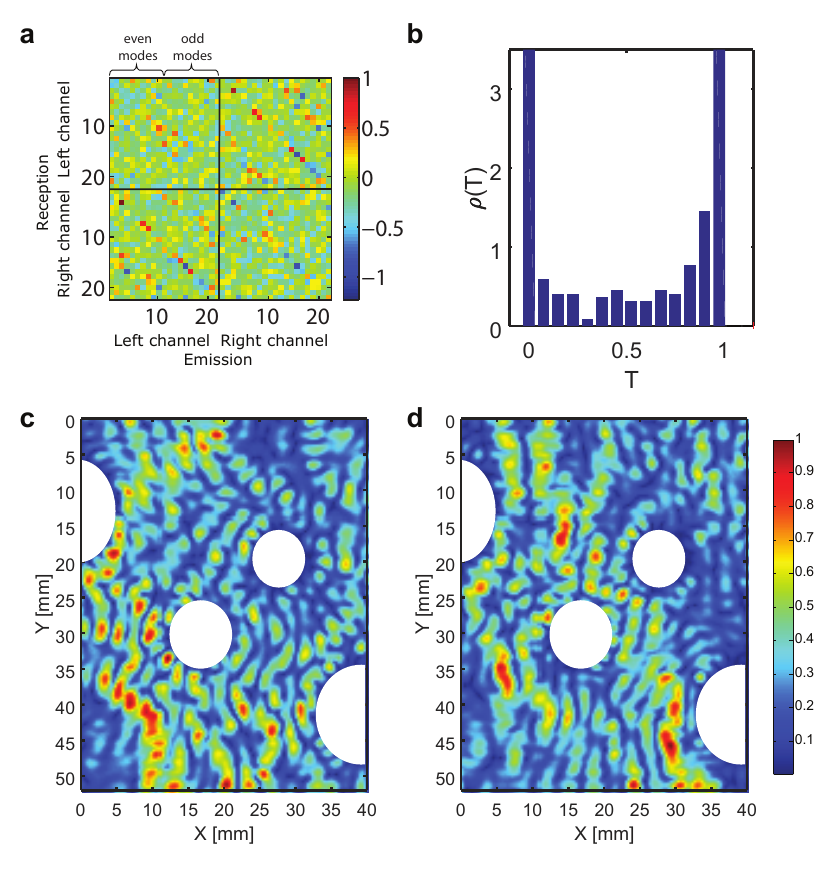}
\caption{({a}), Amplitude of the $\mathbf{S-}$matrix measured in the scattering slab at $f_0 = \unit{0.30}{\mega\hertz}$. The black lines delimit transmission and reflection matrices. ({b}), Transmission eigenvalue histograms, $\rho\left(T\right)$, averaged over the frequency bandwidth. ({c})-({d}), Absolute value of the monochromatic wave-fields associated with two open channels at the central frequency.}
\label{figS1}
\end{figure}

Figure~\ref{figS1}(a) displays an example of $\mathbf{S-}$matrix recorded in the disordered wave guide depicted in Fig.~1(a) of the accompanying paper. Despite a moderate level of disorder, the $\mathbf{S}-$matrix exhibits an overall random appearance. Nevertheless, a residual ballistic wave-front slightly emerges along the diagonal of the transmission matrices. 
The statistics of the transmission eigenvalues $T_l$ computed from the $\mathbf{t-}$matrix is now investigated. As for the regular cavity, their distribution $\rho\left(T\right)$ is estimated by averaging their histograms over the frequency bandwidth. Fig.~\ref{figS1}(b) displays the estimator of this distribution. Most of the transmission coefficients are found to be either close to zero or one, accounting respectively for closed and open channels. Figures~\ref{figS1}(c) and \ref{figS1}(d) display the wave-fields associated with two open channels at the central frequency (see Methods). As in the cavity, the open channels combine multiple path trajectories. As a consequence, they undergo a strong spatial and temporal dispersion while propagating through the scattering slab. As shown in Fig.6 of the accompanying paper, the study of the time-delay matrix allows to lift this degeneracy.

\section{Stability of the particle-like states in the frequency domain}
\begin{figure}[!ht]
\centering
\includegraphics[width=\columnwidth]{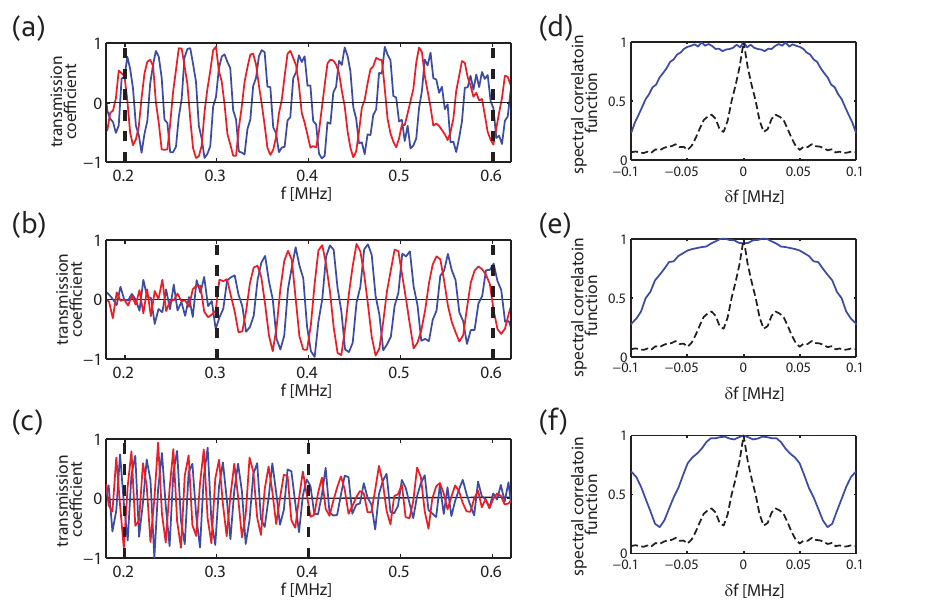}
\caption{\cora{(a)-(b)-(c), Frequency dependence of the transmission coefficients $t^{(q)}_m(f)$ for the three cavity particle-like states displayed in Fig.~4(a)-(b)-(c), respectively (blue line: real part, red line: imaginary part). The black vertical dashed lines delimit the bandwidth of each state. (d)-(e)-(f), Spectral correlation function $|C_m(\delta f) |$ (blue line) of the transmission coefficient $t^{(q)}_m(f)$ displayed in (a)-(b)-(c), respectively. $C_m(\delta f) $ is computed by considering the central frequency $f_0$ of each state. Each curve is compared with the spectral correlation function $C(\delta f) $ of the transmission matrix elements (black dashed line).}}
\label{figS2}
\end{figure}
\cora{Figures~\ref{figS2}(a) and \ref{figS2}(c) display the frequency dependence of the transmission coefficients $t^{(q)}_m(f)$ for the three cavity particle-like states displayed in Fig.~4. The frequency evolution of $t^{(q)}_m(f)$ allows to delimit the frequency range over which each particle-like state remains stable: $f=0.2-0.6$ MHz [Fig.~\ref{figS2}(a)], $f=0.3-0.6$ MHz [Fig.~\ref{figS2}(b)] and $f=0.2-0.4$ MHz [Fig.~\ref{figS2}(c)]. To be more quantitative, we have estimated the spectral correlation function of each state,
$$C_m(\delta f)=t^{(q)}_m(f_0) t^{(q)*}_m(f_0+\delta f), $$
with $f_0$ the associated central frequency. The result is displayed in Figs.~\ref{figS2}(d), \ref{figS2}(e) and \ref{figS2}(f), and compared to the mean spectral correlation function,
$$C(\delta f)=\left \langle t_{ij}(f) t^{*}_{ij}(f+\delta f) \right \rangle, $$
of the measured transmission matrix coefficients $t_{ij}(f)$. The symbol $\langle . \rangle $ here denotes an average over the frequency bandwidth. The FWHM of $C_m(\delta f)$ yields the spectral correlation \cors{width} $\delta f_m$ of each state displayed in Fig.~4. We find $\delta f_m=$ 0.16 MHz [Fig.~\ref{figS2}(d)], 0.17 MHz [Fig.~\ref{figS2}(e)] and 0.12 MHz [Fig.~\ref{figS2}(f)]. These values should be compared to the frequency correlation \cors{width} $\delta f_c=0.02$ MHz of the transmission matrix coefficients. We see that the spectral correlation \cors{width} $\delta f_m$ of particle-like states is 8, 8.5 and 6 times larger than $\delta f_c$. This illustrates the stability of the cavity particle-like states in the frequency domain and accounts for the spatio-temporal focusing of particle-like wave packets in the time domain (see Supplementary Movies 1, 2 and 3).}

\begin{figure}[!ht]
\centering
\includegraphics[width=\columnwidth]{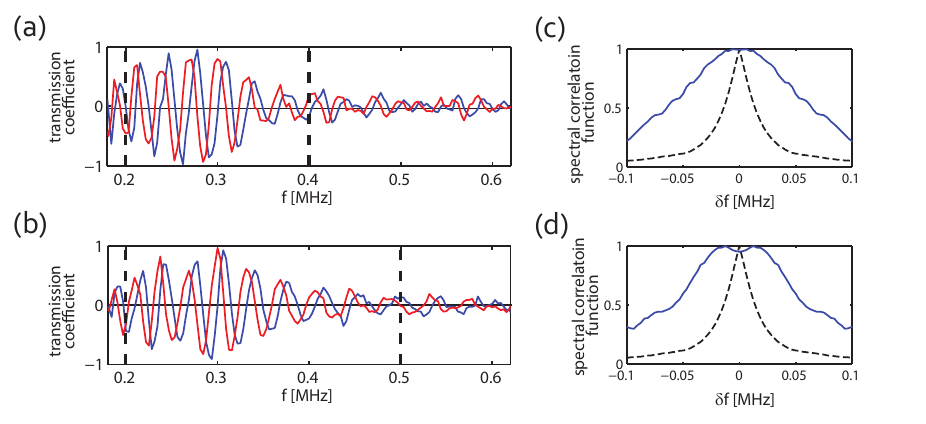}
\caption{\cora{(a)-(b), Frequency dependence of the transmission coefficients $t^{q}_m(f)$ for the two particle-like states in the disordered wave guide, displayed in Figs.~6(c) and 6(d), respectively (blue line: real part, red line: imaginary part). The black vertical dashed lines delimit the bandwidth of each state. (c)-(d), Spectral correlation function, $| C_m(\delta_f) |$, for the transmission coefficient $t^{q}_m(f)$ displayed in (a)-(b), respectively. $C_m(\delta f) $ is computed by considering the central frequency $f_0$ of each state. Each curve is compared with the spectral correlation function $|C(\delta f) |$ of the transmission matrix elements (black dashed line).} }
\label{figS3}
\end{figure}
\cora{The same analysis can be performed for the disordered wave guide. As an example, we investigate the frequency stability of the particle-like states displayed in Figs.6(c) and 6(d). Figures \ref{figS3}(a) and \ref{figS3}(b) display the frequency dependence of the corresponding transmission coefficients $t^{(q)}_m(f)$. Each state remains stable over a frequency range $f=0.2-0.4$ MHz [Fig.\ref{figS3}(a)] and $f=0.2-0.5$ MHz [Fig.\ref{figS3}(b)]. The corresponding spectral correlation functions $C_m(\delta f)$ are displayed in Figs.\ref{figS3}(c) and \ref{figS3}(d). We measure a frequency correlation \cors{width} $\delta f_m$ of 0.13 and 0.12 MHz. Contrastingly, the transmission matrix elements exhibit a frequency correlation \cors{width} $\delta f_c=0.025$ MHz. $\delta f_m$ is thus 5 times higher than $\delta f_c$. This demonstrates the frequency stability of these two particle-like states and accounts for the spatio-temporal focusing of the corresponding particle-like wave packets in the time domain (see Supplementary Movies 4 and 5).}

\section{Captions of the Supplementary Movies}

\noindent \href{http://www.institut-langevin.espci.fr/IMG/avi/movie_1.avi}{Movie 1}: Particle-like wave-packet in the cavity synthesized from the scattering state displayed in Fig. 4(a) over the frequency range $f=0.2-0.6$ MHz.

\noindent \href{http://www.institut-langevin.espci.fr/IMG/avi/movie_2.avi}{Movie 2}: Particle-like wave-packet in the cavity synthesized from the scattering state displayed in Fig. 4(b) over the frequency range $f=0.3-0.6$ MHz.

\noindent \href{http://www.institut-langevin.espci.fr/IMG/avi/movie_3.avi}{Movie 3}: Particle-like wave-packet in the cavity synthesized from the scattering state displayed in Fig. 4(c) over the frequency range $f=0.2-0.4$ MHz.

\noindent \href{http://www.institut-langevin.espci.fr/IMG/avi/movie_4.avi}{Movie 4}: Particle-like wave-packet in the disordered wave guide synthesized from the scattering state displayed in Fig. 6(c) over the frequency range $f=0.2-0.4$ MHz.

\noindent \href{http://www.institut-langevin.espci.fr/IMG/avi/movie_5.avi}{Movie 5}: Particle-like wave-packet in the disordered wave guide synthesized from the scattering state displayed in Fig. 6(d) over the frequency range $f=0.2-0.5$ MHz.

\end{document}